\documentclass[fleqn,usenatbib]{mnras}

\usepackage{newtxtext}
\usepackage[slantedGreek]{newtxmath}

\usepackage[T1]{fontenc}

\DeclareRobustCommand{\VAN}[3]{#2}
\let\VANthebibliography\thebibliography
\def\thebibliography{\DeclareRobustCommand{\VAN}[3]{##3}\VANthebibliography}

\usepackage{graphicx}	
\usepackage{amsmath}	
\usepackage{cleveref}
\usepackage{float}
\usepackage{subcaption}
\usepackage{xcolor}

\usepackage{tabularx}

\title[MAXI J1820$+$070 Complete Outburst]{The Accretion -- Ejection Connection in the Black Hole X-ray Binary MAXI J1820$+$070}

\author[J. S. Bright et al.]{Joe S. Bright,$^{1,2}$\thanks{E-mail: joe.bright@physics.ox.ac.uk} Rob Fender,$^{1,3}$ 
David M. Russell,$^{4}$ 
Sara E. Motta,$^{5}$ 
Ethan Man,$^{1}$\newauthor 
Jakob van den Eijnden,$^{1,6,7}$ 
Kevin Alabarta,$^{4}$ 
Justine Crook-Mansour,$^{1}$ 
Maria C. Baglio,$^{5}$ 
David A. Green,$^{8}$\newauthor 
Ian Heywood,$^{1,2,9,10}$ 
Fraser Lewis,$^{11,12}$ 
Payaswini Saikia,$^{4}$ 
Paul F. Scott,$^{8}$ 
David J. Titterington$^{8}$
\\
$^{1}$Astrophysics, Department of Physics, University of Oxford, Keble Road, Oxford OX1 3RH, UK\\
$^{2}$Breakthrough Listen, Astrophysics, Department of Physics, University of Oxford, Keble Road, Oxford OX1 3RH, UK\\
$^{3}$Department of Astronomy, University of Cape Town, Private Bag X3, Rondebosch 7701, South Africa\\
$^{4}$Center for Astrophysics and Space Science (CASS), New York University Abu Dhabi, PO Box 129188, Abu Dhabi, UAE\\
$^{5}$Istituto Nazionale di Astrofisica, Osservatorio Astronomico di Brera, via E. Bianchi 46, 23807 Merate (LC), Italy\\
$^{6}$Department of Physics, University of Warwick, Coventry, CV4 7AL, UK\\
$^{7}$Anton Pannekoek Institute for Astronomy, Universiteit van Amsterdam, Science Park 904, 1098, XH, Amsterdam, The Netherlands\\
$^{8}$Astrophysics Group, Cavendish Laboratory, University of Cambridge, Cambridge, CB3 0HE, UK\\
$^{9}$Department of Physics and Electronics, Rhodes University, PO Box 94, Makhanda 6140, South Africa\\
$^{10}$South African Radio Astronomy Observatory, 2 Fir Street, Black River Park, Observatory 7925, South Africa\\
$^{11}$Faulkes Telescope Project, Cardiff, Wales, UK\\
$^{12}$The Schools' Observatory, Astrophysics Research Institute, Liverpool John Moores University, 146 Brownlow Hill, Liverpool L3 5RF, UK\\
}

\date{Accepted XXX. Received YYY; in original form ZZZ}

\pubyear{2024}

\begin{document}
\label{firstpage}
\pagerange{\pageref{firstpage}--\pageref{lastpage}}
\maketitle

\begin{abstract}
The black hole X-ray binary MAXI J1820$+$070 began its first recorded outburst in March 2018, and remained an active radio, X-ray, and optical source for over four years. Due to the low distance to the source and its intrinsically high luminosity MAXI J1820$+$070 was observed extensively over this time period, resulting in high-cadence and quasi-simultaneous observations across the electromagnetic spectrum. These data sets provide the opportunity to probe the connection between accretion and the launch of jets in greater detail than for the majority of black hole X-ray binaries. In this work we present radio (Arcminute Microkelvin Imager Large Array, MeerKAT), X-ray (\textit{Swift}), and optical (Las Cumbres Observatory) observations of MAXI J1820$+$070 throughout its entire outburst, including its initial hard state, subsequent soft state, and further hard--state--only re-brightenings (covering March 2018 to August 2022). Due to the regularity and temporal density of our observational data we are able to create a Radio -- X-ray -- Optical activity plane where we find a high degree of correlation between the three wave bands during the hard states, and observe hysteresis as MAXI J1820$+$070 enters and exits the soft state. Based on the morphology of the optical light curves we see evidence for optical jet contributions during the soft--to--hard state transition, as well as fading optical emission well before the hard to soft transition. We establish that the remarkably similar profiles of the re-brightening events are broadly consistent with modified disk instability models where irradiation from the inner accretion disk is included.
\end{abstract}

\begin{keywords}
 stars:black holes, X-rays:binaries, radio continuum:transients
\end{keywords}



\section{Introduction}
Black hole X-ray binaries (BHXRBs), systems containing a stellar-mass black hole accreting material via an accretion disk from a main sequence companion, undergo outbursts during which they cycle through characteristic accretion states \citep[e.g.][]{fender2004,remillard2006}. These accretion states are defined predominantly via X-ray observations, particularly the X-ray hardness ratio and timing properties of a system. The \textit{hard} state is characterised by an X-ray spectrum peaking at $\sim100\,\rm{keV}$ with a strong power-law component, and high levels of X-ray variability \citep{homan2005b}. In the \textit{soft} state the power-law spectral component disappears or is significantly suppressed and the spectrum is dominated by a multi-component black body peaking at $\sim1\,\rm{keV}$, as expected from a multi-temperature accretion disk, and the X-ray variability is significantly suppressed (e.g. \citealt{belloni2010}). During transitions between the hard and soft states BHXRBs are said to be in an intermediate state. Strongly coupled to the accretion state of a BHXRB are its jet properties, observed primarily at radio and (sub-)mm frequencies (e.g. \citealt{fender2004,russell2014,tetarenko2019,russell2020,tetarenko2021}). In the hard state sources emit flat spectrum synchrotron radio emission, consistent with a constantly replenished and collimated outflow (e.g. \citealt{blandford1979,falcke1995}). The presence of this so-called core jet has been directly confirmed through high resolution radio imaging of GRS 1915$+$105 \citep{dhawan2000}, Cygnus X-1 \citep{stirling2001}, and most recently and spectacularly in Swift~J1727.8$-$1613 \citep{wood2024}. In the soft accretion state, emission from this core jet is significantly suppressed, or switches off completely, with the current best constraint on the quenching factors on the order of three dex \citep{russell2020,bright2020,maccarone2020}. 
 
Most BHXRBs spend the majority of their lifetimes accreting at an extremely low rate (e.g,. \citealt{gallo2005,plotkin2013,plotkin2021,carotenuto2022}) in a quiescent state (with an unionised disk) or a very dim hard state \citep{gallo2005}, with their accretion disk (at optical and X-ray wavelengths) and jets (radio wavelengths) frequently undetectable with current observational facilities. Occasionally, likely due to instabilities in the disk, the accretion rate can increase by many orders of magnitude ($\sim8$ dex in the most extreme cases; \citealt{rodriguez2020}) resulting in detectable optical and X-ray emission from the (outer and inner) disk, and the presence of compact core jets detectable in radio. A source undergoing such an instability is said to be in outburst. While some sources fade to quiescence after spending time in the hard-state-only \citep[e.g.][]{tetarenko2016,williams2020,Alabarta2021}, other sources make one or many hard to soft state transitions before returning to the hard state and fading to quiescence \citep[e.g.][]{fender1999,corbel2001,bright2020,carotenuto2021a}. Associated with the hard to soft state transition (but not the reverse one) is the launch of bipolar ejections which can travel to large angular separations from the site of the black hole (when compared to the observed head of the core jet), appear no  longer physically connected to the black hole, and are resolved from the core position at both radio \citep[e.g.][]{mirabel1994,fender1999,bright2020,wood2021,carotenuto2021a,bahramian2023} and X-ray \citep{corbel2002b,migliori2017,espinasse2020} frequencies. 

The BHXRB MAXI J1820$+$070 (hereafter J1820) has been the target of intense monitoring campaigns across the electromagnetic spectrum since it began its first known outburst in March 2018, with its brightness allowing observers to collect a wealth of high quality data from the source and progress our understanding of accretion and jet launching in BHXRBs \citep[e.g.][]{shidatsu2019,bright2020,shaw2021,trujillo2024}. J1820 was discovered as an optical transient by the All-Sky Automated Survey for SuperNovae (ASAS-SN; \citealt{shappee2014, kochanek2017}) on 06 March 2018 and labelled ASASSN-18ey \citep{tucker2018}. Initially classified as a cataclysmic variable (due to its apparent association with a Gaia source), J1820 was then detected by the Monitor of All-sky X-ray Image (MAXI; \citealt{matsuoka2009}) Gas Slit Camera \citep{mihara2011,sugizaki2011}, assigned the name MAXI J1820, and associated with ASASSN-18ey on 11 March 2018 \citep{MAXI_discovery, MAXI_ASASSN_association} where initial observational evidence suggested it was a BHXRB \citep{BHXRB_classification}. J1820 initially underwent a `classical' outburst, with a single hard to soft state transition, followed by a return to the hard state. Instead of immediately returning to quiescence J1820 then underwent three hard-state-only re-brightenings reaching 1 to 10 per cent of the peak luminosity of the initial outburst depending on observing band. The intensive monitoring of J1820 has lead to a number of exciting results. These include an accurately measured radio parallax distance of $2.96\pm0.33$ kpc and position RA$=18^{\rm h}20^{\rm m}21\fs938$, Dec$=07\degr11\arcmin07\farcs165$ \citep[along with a well constrained proper motion][]{atri2020}, the detection of large scale jet ejections during the hard to soft state transition and strong constraints on their energetics \citep{bright2020,espinasse2020,wood2021}, an association between the launch of these transient ejecta and the properties of the accretion flow \citep{homan2020,ma2021,bei2021,wang2021,ma2023}, measured changes in the geometry of the corona \citep{kara2019,ma2023}, a dynamical confirmation that J1820 harbours a black hole with a low mass companion \citep{torres2019,torres2020}, and the ability to track the spectral break of the hard state jet over three decades in frequency \citep{trujillo2024}, to name but a few.

BHXRBs are ideal systems for probing the connection between accretion and the production of jets owing to their rapid evolutionary timescales during outburst. While it is common to correlate X-ray observations of BHXRBs, which probe the inner accretion disk, with radio observations, which probe the core jet, optical observations are considered in tandem less often (although see e.g. \citealt{russell2006}). The origin of the optical emission in BHXRBs is more ambiguous than that of the radio and X-rays, with direct accretion disk photons \citep{shakura1973}, reprocessed accretion disk photons \citep[e.g.][]{cunningham1997,vrtilek1990}, and jet photons \citep[e.g.][]{corbel2002a,chaty2003,homan2005a}, or some combination of the three, all possible contributing sources (see e.g. \citealt{russell2006,veledina2013,vincentelli2021,saikia2022,saikia2023b,koljonen2023} for additional interpretations). Additionally, the contribution of the companion star will dominate in the optical and infrared, although this is more prominent for XRBs harbouring a high mass companion (and less so at the peak of outburst compared to quiescence). Due to this ambiguity, only when high quality multi-wavelength data exists, as is the case for J1820, can we attempt to disentangle these optical emission mechanisms and put them in context with the more frequently produced radio -- X-ray correlation. Such observations provide insight into the emission profile and evolution of jets, how their formation is coupled to the inner accretion disk, and how different regions of the accretion disk interact. 

In this paper we present observations of J1820 at radio, optical, and X-ray frequencies between MJD 58189 (12 March 2018) and MJD 59792 (1 August 2022). In \Cref{section:observations} we describe our observations of J1820 and relevant data reduction techniques, in \Cref{section:results} we present our radio, optical and X-ray light curves and correlations between them, and in \Cref{sec:discussion,section:conclusions} we discuss the light curves and correlations in the context of BHXRB accretion models.

\section{Observations}\label{section:observations}

\subsection{Radio}
\subsubsection{Arcminute Microkelvin Imager Large Array}
Observations of MAXI J1820 with the Arcminute Microkelvin Imager Large Array (AMI-LA; \citealt{zwart2008,hickish2018}) were triggered automatically in response to a SWIFT Burst Alert Telescope VOEvent (trigger ID 813771) on 12 March 2018 \citep{staley2013, BAT_discovery_ATel}. Manual observations were regularly scheduled thereafter, and remain ongoing. Data were recorded at a central frequency of $15.5\,\rm{GHz}$ and across a $5\,\rm{GHz}$ bandwidth consisting of 4096 frequency channels, which we average to eight channels before processing. The AMI-LA data, which measures Stokes $I+Q$, were reduced using a custom pipeline, \textsc{reduce\_dc}, which flags data for instrumental issues, antenna shadowing, the effects of weather conditions, and settling times. The software then applies bandpass, absolute flux scale (using either 3C48 or 3C286), and time dependent phase correction \citep[see e.g.][]{davies2009,perrott2013}. The compact source J1824$+$1044 was used as the phase reference calibrator for all but the first (automatically triggered) observation, which used J1819$+$0640. We then perform additional flagging using the Common Astronomy Software Applications (CASA 6.4.1.12; \citealt{mcmullin2007,casa2022}) package, using the \textsc{rflag} and \textsc{tfcrop} flagging modes. Imaging is performed with the \textsc{tclean} task, where we use a Briggs \citep{briggs1995} robust weighting of 0.5, which we find to be a good compromise between sensitivity and dirty beam quality, and two Taylor terms to account for the high fractional bandwidth when cleaning. The cell size was fixed to $5$ arcsec and we produced $512 \times 512$ pixel$^2$ images which cover the primary beam of the AMI-LA while appropriately sampling the synthesised beam which has typical major and minor axis of 30 and 60 arcsec, respectively (assuming a Gaussian beam). To measure the flux density in each observation we use the CASA task \textsc{imfit}, forcing the Gaussian source component to have the same shape as the synthesised beam of the particular observation. Due to the brightness of MAXI J1820 in its initial outburst, as well as the re-brightenings, we are able to measure the amplitudes directly from the complex visibilities, and observe the short timescale (minutes) variability of J1820, which revealed a flare during the initial hard to soft state transition \citep{bright2020}. No other such short term variability is seen during our observations with the AMI-LA. The initial automatically triggered observation was calibrated and imaged manually and the peak pixel was taken as the flux density, with a conservative error of 20 per cent.

\begin{figure}
    \centering
    \includegraphics[width=\columnwidth]{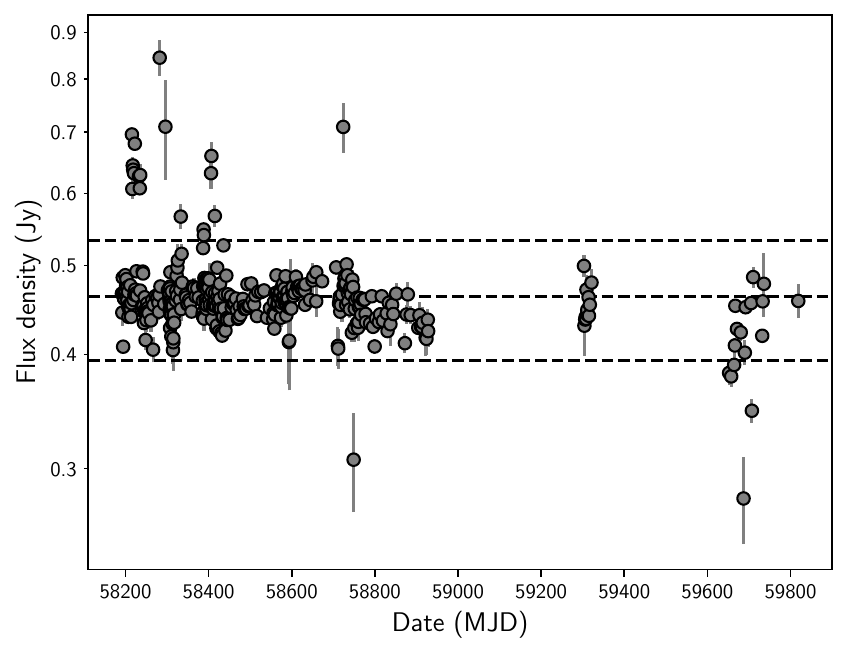}
    \caption{The light curve of the phase reference calibrator J1824$+$1044. Errors on data points indicate the statistical uncertainty from the \textsc{imfit} task. The central dashed line marks the mean flux density of J1824$+$1044 and the lines above and below it mark 1.15 times the mean flux density and 0.85 times the mean flux density, respectively. We reject observations of J1820 for which the flux density of J1824$+$1044 is outside of this range as they indicate a poor transfer of the flux density scale for that observation.}
    \label{fig:J1824_lc}
\end{figure}

In this work we began with a total of 396 distinct observations of J1820 (defined as individually scheduled observations in the AMI-LA control system, including the initial automatically triggered observation). This large number of observations motivated the automated reduction and imaging procedure described above, however we perform a number of quality cuts on our data before presenting the final light curve of J1820 from the AMI-LA. Firstly, using the CASA task \textsc{imfit}, we create a light curve of the phase reference calibrator by fitting a point source component to J1824$+$1044 in the image plane and rejecting observations (both the calibrator and J1820) for which either the fit did not converge or the ratio of the major to minor full width half maximum of the synthesised beam was greater than 10 (which indicates antenna dropout). This cut resulted in the rejection of 11 observations. Phase reference calibrators should be point like and non-variable on the timescale of an observation length. Longer term variability in the flux density of a phase reference calibrator is acceptable, but apparently large short timescale variability could also indicate poor calibration of the flux density scale for a given observation. As such, we calculate the average value of the flux density measured from the phase reference calibrator and reject observations (both the calibrator and J1820) where we measure the flux density of the phase calibrator to be discrepant from the average by more than 15 per cent, which resulted in a further 24 observations being rejected. The light curve of J1824$+$1044 is shown in \Cref{fig:J1824_lc} and demonstrates a number of flux density measurements significantly different from the average, which would imply large flux density changes on short (day length) timescales. This process filtered out obvious outliers in the light curve of J1820. After removing these 35 observations the calibrator is variable at the $\sim5$ per cent level whereas before filtering out the 24 unreliable measurements the variability was $\sim10$ per cent. Based on this we assign a 5 per cent uncertainty to our flux measurements of J1820 added in quadrature to the statistical uncertainty derived from the image plane fitting. Finally, we manually inspected all images of J1820 for which the flux density was below $500\,\upmu\rm{Jy}$, resulting in the rejection of 34 additional observations. Observations for which the fitted flux density was less than five times the image RMS noise were labelled as upper limits. After these cuts, we are left with 326 distinct observations (including upper limits). We include a sample of our AMI-LA observations in \Cref{tab:ami_data} and a machine readable table covering our entire observing campaign is available as part of the online version of this article. 

\subsubsection{MeerKAT}
As part of the ThunderKAT large survey project \citep{fender2016} we began observing the field of MAXI J1820 with the MeerKAT telescope on 28 September 2018 when the source appeared clearly extended (with angular extents $\sim10$ arcsec) due to the presence of large scale ejections launched during the hard to soft state transition \citep{bright2020}, with a typical MeerKAT resolving beam being $\sim7$ arcsec. Note that this does not violate our assumption of a point source for our AMI-LA observations as during the soft state the transient ejecta were significantly closer to the core, and are subdominant at $15.5\,\rm{GHz}$ once the hard state core jet reignites. We continued observing J1820 with MeerKAT until August 2022. MeerKAT data were reduced using the \textsc{oxkat} data reduction pipeline \citep{heywood2020, kenyon2018, makhathini2018, hugo2022, offringa2014} which performs phase reference calibration as well as phase and amplitude self-calibration (we do not employ direction dependent calibration techniques, although \textsc{oxkat} includes that capability, due to the relative simplicity of the sky brightness distribution in the region surrounding MAXI J1820). We used the flux standard J1939$-$6342 to calibrate the absolute flux density scale and bandpass response of the instrument, while J1733$-$1304 was used to calibrate the time-dependent complex gains. We used a Briggs robust weighting of $-0.3$ when imaging \citep{briggs1995}.

For observations taken on and before MJD 58454 core flux densities are taken from \citet{bright2020}, where multiple point source components were used to describe the flux of the core and approaching and receding transient ejecta. The same data reduction procedure was used in \citet{bright2020} as for this work. For observations between MJD 58460 and MJD 58551 there is still low-level emission from the receding ejections component but we are unable to satisfactorily fit multiple emission components to the extended morphology. We fit an unresolved point source to these data as the core is the dominant component but note that the low level emission might distort the fit to a small extent. After MJD 58511 the emission from J1820 is entirely core dominated, and we fit an unresolved source to a small region around the source position. A 5 per cent calibration uncertainty is added in quadrate to the uncertainty derived from \textsc{imfit}. The MeerKAT data detailing the flux density of the core jet from J1820 are (partially) given in Table \ref{tab:meerkat_data}, with a full machine readable table covering our entire observing campaign available as part of the online version of this article.

\begin{table}
\centering
\caption{Arcminute Microkelvin Imager Large Array observations of MAXI J1820. A full machine readable table is available as part of the online version of this article. Errors are statistical only, and should be combined in quadrature with a 5 per cent calibration uncertainty. Note that the first observation was triggered automatically in response to a \textit{Swift}-BAT VOEvent trigger.}
\label{tab:ami_data}
\begin{tabular}{ccc}
\hline
Date & MJD midpoint & Flux density\\
$[$dd/mm/yy$]$ & & [mJy]\\
\hline
12/03/18 & 58189.1121 & $2.49\pm0.50$ \\
14/03/18 & 58191.2914 & $17.13\pm0.16$ \\
15/03/18 & 58192.2887 & $21.59\pm0.45$ \\
16/03/18 & 58193.2860 & $31.69\pm0.25$ \\
18/03/18 & 58195.2805 & $50.78\pm0.27$ \\
... & ... & ... \\
\hline
\end{tabular}
\end{table}

\begin{table}
\centering
\caption{Meerkat observations of MAXI J1820. Errors are statistical only and as such a 5 per cent calibration error should be added in quadrature when using these data. A full machine readable table is available as part of the online version of this article. The `Obs. ID' column can be used to identify observing blocks in the MeerKAT data archive (\url{https://apps.sarao.ac.za/katpaws/archive-search}).}
\label{tab:meerkat_data}
\begin{tabular}{cccccc}
\hline
Obs. ID & Date & MJD & Flux & Image \\
 &  & midpoint & density & RMS\\
& [dd/mm/yy] & & [mJy] & [$\upmu\rm{Jy}\,\rm{beam}^{-1}$]\\
\hline
1538156623 & 28/09/18 & 58389.75 & $3.47\pm0.05$ & 41\\
1538757039 & 05/10/18 & 58396.70 & $11.8\pm 0.1$ & 72\\
1539354654 & 12/10/18 & 58403.66 & $2.62\pm0.04$ & 37\\
1539529257 & 14/10/18 & 58405.67 & $2.41\pm0.03$ & 24\\
1539955889 & 19/10/18 & 58410.62 & $1.52\pm0.06$ & 50\\
... & ... & ... & ... &  ...\\
\hline
\end{tabular}
\end{table}

\subsection{X-rays: \textit{Swift}}
MAXI J1820 was observed thoroughly by the \textit{Swift} X-ray Telescope (hereafter XRT; \citealt{gehrels2004,burrows2005}) on board the Neil Gehrels Swift Observatory (\textit{Swift}) beginning on MJD 58189 (12 March 2018) and ending on MJD 59735 (5 June 2022). Observations were conducted both in Proportional Counter (PC) mode and Windowed Timing (WT) mode. To produce an X-ray light curve for J1820 we first built spectra for each observation ID (obsID) using the \textit{Swift}-XRT data product generator\footnote{\url{https://www.swift.ac.uk/user_objects/}} \citep{evans2009}. Spectra that contained at least 300 total counts were then binned using \textsc{grppha} such that each bin had a minimum of 20 counts. For spectra with less than 300 total counts no binning was performed. To determine the flux associated with each spectrum they were loaded into XSPEC \citep[version: 12.14.0b;][]{arnaud1996} where we ignore bad data and ignore data outside of the range 0.5 to $10\,\rm{keV}$. We discard spectra for which the time on source was less than one minute, there were fewer than 15 counts in WT mode, or there were fewer than 3 counts in PC mode. We then proceed to fit the spectra, with the method depending on the total counts. For spectra with less than 300 total counts we use Cash statistics to constrain our fitting and only consider an absorbed power-law model (\textsc{tabs(pow)} in XSPEC), with fixed neutral hydrogen column density ($n_{\rm H}=0.091$), and fixed photon index ($\Gamma=1.7$) if the total counts were below 50. After identifying the best fit model parameters we freeze them and introduce a flux parameter so that the new model is \textsc{tabs*cflux(pow)} and \textsc{cflux} is the only free parameter. This model is then fit again to constrain \textsc{cflux} and its value and one sigma error are extracted. For spectra with 300 or above total counts the procedure is much the same, however during the soft and intermediate states defined in \citep{shidatsu2019} we compare both a power law and power law plus disk blackbody (\textsc{tabs(pow + diskbb)} in XSPEC) model, selecting the favoured model using an F-test with a $p$ value below 0.001 required to accept the more complex model. As in the previous case we then calculated the final flux through the addition of the \textsc{cflux} parameter. Due to the higher counts for these spectra we use the Chi-squared test statistic to determine the quality of the fit. A sample of our X-ray observations are given in \Cref{tab:xrt_data} with a full version available in the online version of this article.

\begin{table}
\centering
\caption{\textit{Swift}-XRT observations of MAXI J1820. A full machine readable table is available as part of the online version of this article.}
\label{tab:xrt_data}
\begin{tabular}{ccccc}
\hline
Mode & Date & MJD midpoint & Flux\\
& [dd/mm/yy] & & [$\times10^{-9}\,\rm{erg}\,\rm{s}^{-1}$]\\
\hline
WT & 12/03/18 & 58189.0814 & $1.91\pm0.01$\\
PC & 12/03/18 & 58189.1217 & $1.52\pm0.01$\\
WT & 13/03/18 & 58190.7401 & $4.19\pm0.02$\\
WT & 14/03/18 & 58191.8369 & $6.10\pm0.01$\\
WT & 14/03/18 & 58191.8786 & $5.86\pm0.02$\\
... & ... & ... &  ...\\
\hline
\end{tabular}
\end{table}

\subsection{Optical}
Comprehensive optical monitoring of MAXI J1820 was performed with telescopes at the Las Cumbres Observatory (LCO). Data were taken during the initial 2018 outburst, and during the multiple re-brightening events \citep[e.g.][]{BHXRB_classification,Baglio2021ATel14492,Russell2019XBNEWS,2019ATel12534}, as part of a monitoring campaign of $\sim$50 low-mass X-ray binaries coordinated with the Faulkes Telescope Project \citep{Lewis2008,Lewis2018}. The monitoring includes data taken at the 2-m Faulkes Telescopes at Haleakala Observatory (Maui, Hawai`i, USA) and Siding Spring Observatory (Australia), and the 1-m telescopes at Siding Spring Observatory, Cerro Tololo Inter-American Observatory (Chile), the South African Astronomical Observatory (SAAO; South Africa), and the McDonald Observatory (Texas, USA). For the purposes of this work we make use of the images taken using the SDSS $g^{\prime}$ and $i^{\prime}$ filters.

The LCO data were initially processed using the LCO Banzai pipeline \citep{LCO_banzai}. Reduction and analysis of the reduced images was achieved using the real-time data analysis pipeline, \textsc{XB-NEWS} \citep[the X-ray Binary New Early Warning System; see][]{Russell2019XBNEWS,Pirbhoy2020,Goodwin2020}. The \textsc{XB-NEWS} pipeline downloads images of targets of interest from the LCO archive soon after they are taken by the telescopes, along with their associated calibration data. Then, the pipeline performs quality control steps to ensure that only good quality images are analysed, and produces an astrometric solution for each image using Gaia DR2 positions\footnote{\url{https://www.cosmos.esa.int/web/gaia/dr2}}. Aperture photometry was then performed on all the stars in each image, solving for the zero-point calibrations between epochs \citep{Bramich2012}. The ATLAS All-Sky Stellar Reference Catalog \citep[ATLAS-REFCAT2;][]{Tonry2018} was used for flux calibration. The pipeline also performed multi-aperture photometry \citep[azimuthally-averaged PSF profile fitting photometry,][]{Stetson1990} for point sources. We detect the source with high significance throughout the outburst and re-brightenings. We add a systematic 0.04 magnitude error in quadrature with the error reported via the data reduction pipeline. A sample of our optical observations are given in \Cref{tab:optical_data_g,tab:optical_data_i} with a full version available in the online version of this article. We convert our optical magnitudes to luminosities using $L_{O}=4\pi d^{2}10^{-(m + 48.585 - yA_{\nu})/2.5}(c/\lambda)$ where $\rm{m}$ is the apparent optical magnitude, $d$ is the distance to J1820, $c$ is the speed of light, and $\lambda$ is the central frequency of the filter being used. We have corrected for extinction using $A_{\nu}=0.558$ from \citet{tucker2018} and filter specific extinction coefficients ($y$) from \citet{cardelli1989}. All observational data can also be found in machine readable format at the Zenodo DOI 10.5281/zenodo.15721448.

\begin{table}
\centering
\caption{LCO observations of MAXI J1820 taken with the SDSS $g^{\prime}$ filter. The error reported here includes a 0.04 magnitude error combined in quadrature with the statistical error. A full machine readable table is available as part of the online version of this article.}
\label{tab:optical_data_g}
\begin{tabular}{cccc}
\hline
Date & MJD midpoint & Apparent magnitude\\
$[$dd/mm/yy$]$ & & \\
\hline
13/03/18 & 58190.3864 &  $13.11\pm0.04$\\
14/03/18 & 58191.4074 &  $12.93\pm0.04$\\
14/03/18 & 58191.7960 &  $12.90\pm0.04$\\
17/03/18 & 58194.3964 &  $12.42\pm0.04$\\
18/03/18 & 58195.3420 &  $12.43\pm0.04$\\
... & ... &  ...\\
\hline
\end{tabular}
\end{table}

\begin{table}
\centering
\caption{LCO observations of MAXI J1820 taken with the SDSS $i^{\prime}$ filter. The error reported here includes a 0.04 magnitude error combined in quadrature with the statistical error. A full machine readable table is available as part of the online version of this article.}
\label{tab:optical_data_i}
\begin{tabular}{cccc}
\hline
Date & MJD midpoint & Apparent magnitude\\
$[$dd/mm/yy$]$ & & \\
\hline
13/03/18 & 58190.3864 &  $13.04\pm0.04$\\
14/03/18 & 58191.4074 &  $12.93\pm0.04$\\
14/03/18 & 58191.7960 &  $12.84\pm0.04$\\
17/03/18 & 58194.3964 &  $12.52\pm0.04$\\
18/03/18 & 58195.3420 &  $12.17\pm0.04$\\
... & ... &  ...\\
\hline
\end{tabular}
\end{table}

\begin{figure*}
    \centering
    \includegraphics[width=\textwidth]{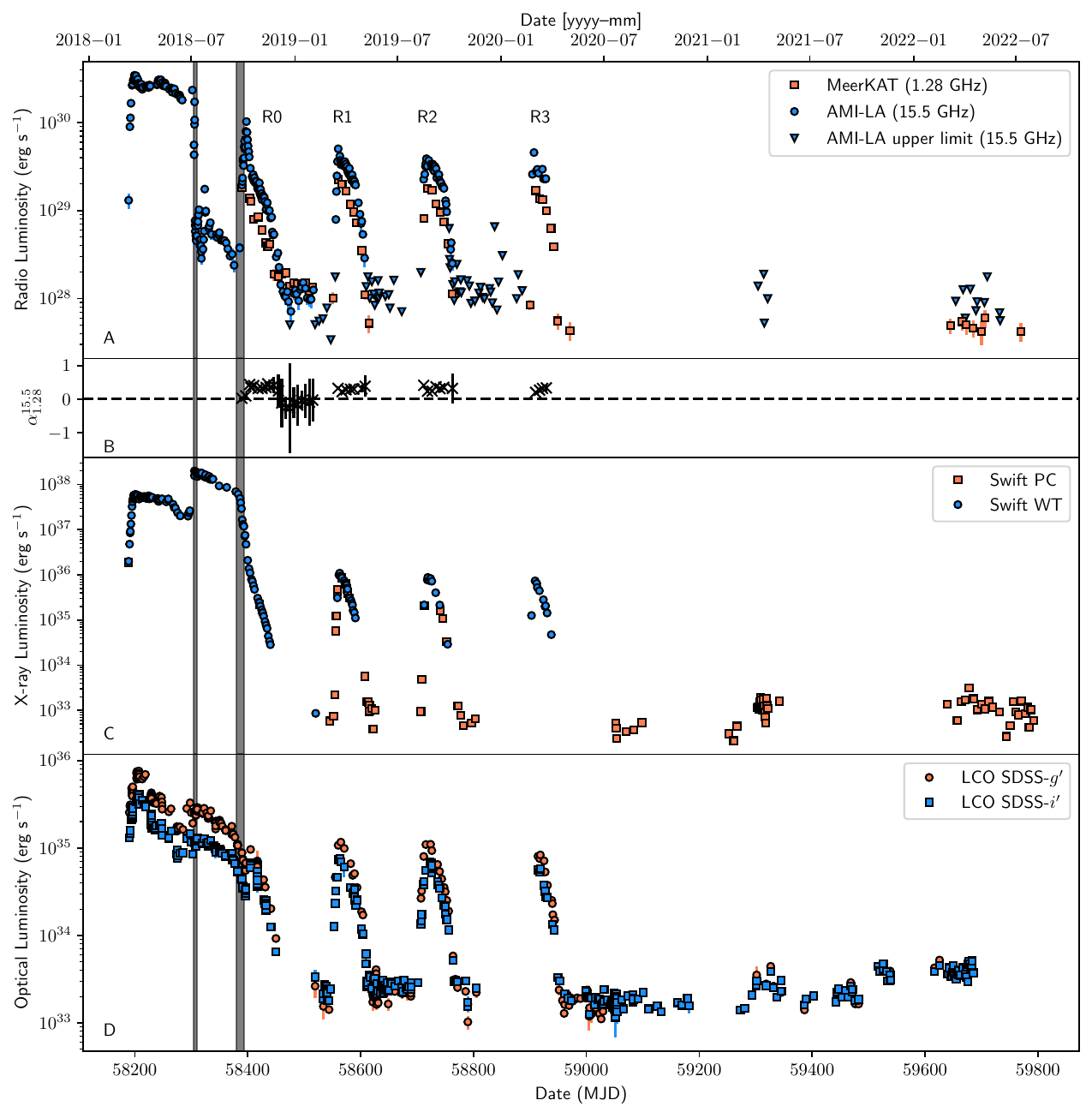}
    \caption{Radio (Arcminute Microkelvin Imager Large Array, MeerKAT), X-ray (\textit{Swift}), and optical (Las Cumbres Observatory) light curves of MAXI J1820. A distance of $2.96\,\rm{kpc}$ was assumed for the conversion to luminosity \citep{atri2020}. Grey shaded regions mark times during which J1820 was in the intermediate state according to \citet{shidatsu2019}. \textbf{Panel A:} MeerKAT $1.28\,\rm{GHz}$ and AMI-LA $15.5\,\rm{GHz}$ radio observations of J1820$+$070. See main text for how flux densities and upper limits were derived. AMI-LA upper limits are 5 sigma. Note that we assume a flat spectrum and scale to a common frequency of $5\,\rm{GHz}$ when converting from flux density to luminosity. \textbf{Panel B:} The Radio spectral index between $1.28\,\rm{GHz}$ and $15.5\,\rm{GHz}$ with the convention $F_{\nu}(\nu)\propto\nu^{\alpha}$. The horizontal dashed black line corresponds to a flat radio spectrum ($\alpha=0$). The spectral index was calculated for each MeerKAT point using the nearest in time AMI-LA observation, if an observations existed within 2 days of the MeerKAT one. \textbf{Panel C:} \textit{Swift} X-ray Telescope $0.5$--$10\,\rm{keV}$ luminosity of J1820 taken both in windowed timing and proportional counter mode. \textbf{Panel D:} Optical $g'$ and $i'$ luminosity of J1820 taken from the Las Cumbres Observatory. The central frequency of the $g'$ and $i'$ filters were used to convert to a luminosity.}
    \label{fig:mw_light_curves}
\end{figure*}

\section{Results}\label{section:results}
\subsection{Light Curves}
\Cref{fig:mw_light_curves} shows the radio, X-ray, and optical luminosity of J1820 spanning $\sim1600$ days since its initial rise out of quiescence. Combinations of the radio, X-ray, and optical observations of the initial hard state, subsequent soft state and the following hard state have been analysed to an extent in other works \citep[e.g.][]{shidatsu2019,shaw2021,tetarenko2021,trujillo2024} but have not been presented together and with such comprehensive coverage as in \Cref{fig:mw_light_curves}. 

In addition to the initial `canonical' outburst, we show data from three hard-state-only (evidenced by the radio spectral index also shown in \Cref{fig:mw_light_curves}) re-brightenings, which we label R1, R2, and R3, in chronological order. These re-brightenings are obvious at radio, X-ray, and optical wavelengths, and show remarkably similar light curve morphologies with a fast rise and a decay characterised by two distinct decay rates, an initially shallow decay followed by a faster decline most easily seen in the AMI-LA $15.5\,\rm{GHz}$ radio data (see also \Cref{fig:re-brightenings} for isolated light curves of the re-brightenings). As expected for hard-state-only outbursts we see no evidence for the launch of transient ejecta in our MeerKAT images during any of the three re-brightenings, although MeerKAT only probes structures down to a size scale of a few seconds of arc. Additionally, it is worth noting the similarity of the hard state immediately following the soft to hard state transition (sometimes called the fading hard state, which we label R0) with the subsequent re-brightenings, minus the initial flare as the core jet reignites. We discuss these similarities in \Cref{sec:discussion_morph}.

After around MJD 59000 J1820 ceased entering into hard-state-only re-brightenings. The lack of regular sampling after this date (partly due to the Coronavirus pandemic in 2019) does not indicate that re-brightening events were missed, as MAXI would have detected such events and AMI-LA/MeerKAT/\textit{Swift} observations would have been triggered in response. Dense optical monitoring was undertaken between MJD $\sim59000$ and MJD $\sim59800$ and showed no re-brightening events in this time. The source was clearly still detected, and appeared to steadily rise in luminosity in both of our optical observing bands significantly above the Panoramic Survey Telescope and Rapid Response System (PanSTARRS) archival limits from the system. True quiescence was only reached as recently as June 2023 (MJD 60106) well beyond the coverage of the observations presented here \citep[see][]{baglio2023atel} indicating a low level of activity after the re-brightening events ended.

\subsection{Correlations}\label{sec:correlations}
Due to our dense temporal sampling of J1820 at radio, X-ray, and optical wavelengths we are able to construct various two way correlations, as well as a radio -- X-ray -- optical activity plane, for the entire outburst. We are also able to produce a correlation for the canonical outburst and re-brightening events separately to check for any differences in behaviour. In all following cases we correlate the data by taking the two (or three) light curves being correlated and dividing the time range covered into bin sizes of 1.5 days (1.8 days for the radio -- X-ray -- optical correlation). These timescales were selected to maximise the number of data points for correlation, without correlating data on timescales longer than those typically seen in X-ray binary evolution. Within each bin we check for observations taken at the wavelengths being correlated. If none are found, that bin is ignored. If a single measurement from each wavelength is found then this is used in the correlation. If multiple measurements are found within a bin then their average is used, with the error on the average taken to be the standard deviation of the measurements within the bin. Upper limits are not considered in any of our correlations.

When fitting two way correlations we use the \textsc{python} \textsc{scipy} orthogonal distance regression (ODR) module to fit power law functions to the correlated data of the form $L_{1}=AL_{2}^{a}$, where $L_{1}$ and $L_{2}$ are luminosities at observing bands 1 and 2, respectively, and we refer to $a$ and $A$ as the power law index and scaling factor, respectively. ODR in \textsc{scipy} accounts for errors on both the `independent' and `dependent' variables ($x$ and $y$) as is appropriate when correlating measured data. We fit the logarithm of our data such that the function describing the correlation is linear. 

When considering the correlation between radio, optical, and X-ray measurements we fit a line to the logarithmic data using a power iteration algorithm which converges to a line of best fit (defined by the centroid, $[L_{\rm{X},0}, L_{\rm{R},0}, L_{\rm{O},0}]$ and unit direction vector $[a, b, c]$) according to a minimisation of the sum of the orthogonal distance in three dimensions. In order to calculate the error on our fitted parameters we assume that the errors on each point in the three dimensional space is Gaussian and we sample 1000 sets of points and perform the fitting for each set, taking the error on each parameter as the standard deviation of the output from the 1000 fitting runs. We give the correlation parametrically as

\begin{gather}
    \begin{bmatrix}L_{\rm{X}}\\L_{\rm{R}}\\L_{\rm{O}}\end{bmatrix} = \begin{bmatrix}L_{\rm{X},0}\\L_{\rm{R},0}\\L_{\rm{O},0}\end{bmatrix} + t\begin{bmatrix}a\\b\\c\end{bmatrix}
\end{gather}

\noindent where $t$ can be used to describe any point on the line.

\subsubsection{Radio -- X-ray Correlation}\label{subsec:rxcorr}
\begin{figure}
\includegraphics[width=\columnwidth]{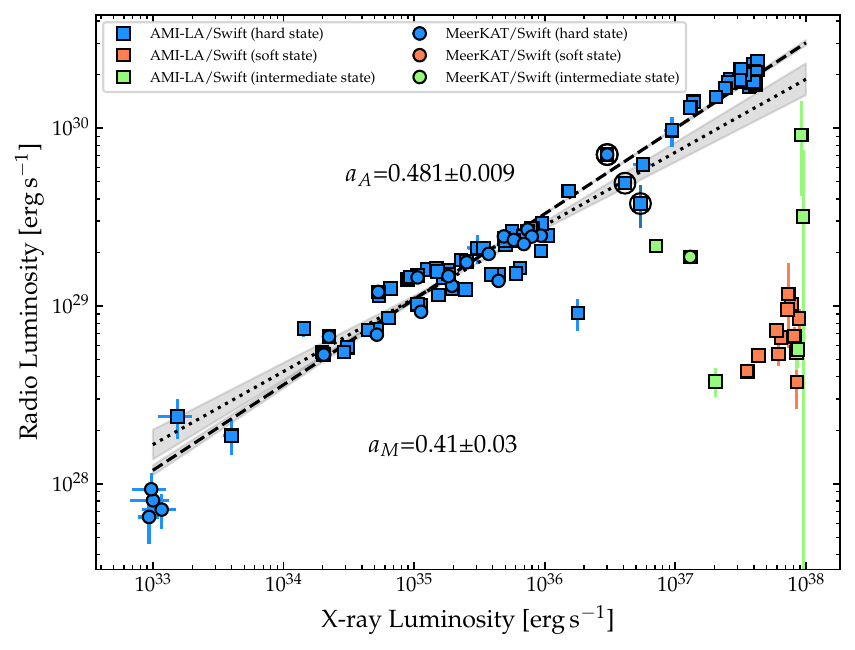}
\caption{The radio -- X-ray correlation for J1820, including data from the initial outburst (as seen in \citealt{bright2020}), and the three re-brightening events shown in \Cref{fig:mw_light_curves}. Squares show data taken with the AMI-LA recorded at $15.5\,\rm{GHz}$ and circles show data from MeerKAT recorded at $1.28\,\rm{GHz}$. Blue, green, and orange indicate the spectral state of the source when the data were recorded, showing hard, intermediate, and soft, respectively. The two annotated dashed lines show best power law fit to the data from the AMI-LA (top line) and MeerKAT (bottom line) with the error region shaded in grey. Soft state data is not included in the fit as it can be attributed to emission from transient ejecta launched during the hard to soft state transition \citep{bright2020}. The data have been scaled to a common luminosity at $5\,\rm{GHz}$ assuming a spectral index as defined in \Cref{subsec:rxcorr}. Hard state observations within 5 days of the state transitions outlined in \citet{shidatsu2019} are encircled.}
\label{figure:radio_xray_correlation}
\end{figure}

During the hard accretion state the radio emission from BHXRBs, emanating from a compact core jet, is strongly coupled to the X-ray emission. To first order this radio -- X-ray correlation can be separated into two `tracks', named radio-loud and radio-quiet (or the standard and outlier tracks, respectively; e.g. \citealt{gallo2012,motta2018}). The majority of BHXRBs lie on the radio-quiet (outlier) track, showing a deviation from the standard track at high luminosities (showing a steeper correlation of $L_{\rm{R}}\propto L_{\rm{X}}^{>1}$) and rejoining it when decaying into quiescence \citep[e.g,][]{coriat2011, carotenuto2021b}. J1820 is an example of a radio-loud source, exhibiting an approximate correlation $L_{\rm{R}}\propto L_{\rm{X}}^{0.6}$ (e.g. \citealt{corbel2000, corbel2003}) that persists over a large range in radio and X-ray luminosities \citep{tremou2020,shaw2021}. During its initial outburst J1820 was shown to follow the radio-loud track with a marginally shallower correlation than expected, with index $0.50\pm0.09$ \citep{bright2020}. The correlation appeared particularly shallow during the hard state immediately following the soft state associated with the launch of transient ejecta (which we label R0), with an index of $0.37\pm0.03$ in the X-ray luminosity region between $\sim10^{34}$ and $\sim10^{36}\,\rm{erg}\,\rm{s}^{-1}$ as is evident in \Cref{figure:radio_xray_correlation} (and can be seen more clearly in figure 1 of \citealt{bright2020}). Combining all of our hard state data we derive a global relation of
\begin{equation}
L_{\rm{R}}\propto L_{\rm{X}}^{0.481\pm0.009}
\end{equation}
for the AMI-LA data, and a marginally shallower
\begin{equation}
L_{\rm{R}}\propto L_{\rm{X}}^{0.41\pm0.03}
\end{equation}
for the MeerKAT data. These are also both consistent with what was derived in \citet{bright2020} and \citet{shaw2021}, with marginally shallower than the canonical value of $\sim0.6$ for radio loud BHXRBs. To convert our flux densities to luminosities we have used the spectral indices in panel B of \Cref{fig:J1824_lc}. If a radio observation is within three days of an epoch where a spectral index is measured, then that spectral index is used to shift the AMI-LA or MeerKAT observation to a flux density at $5\,\rm{GHz}$. If no spectral index is found within three days then the median spectral index (excluding the time range where the spectral index was negative) was used to scale the data. Soft state, intermediate state, and the data where the spectral index was negative were not scaled. These data are not included in the correlations. This process is used for all correlations involving radio observations, but not for light curves, where a flat spectrum at the central observing frequency is assumed. When fitting the correlation we ignore hard state data within 5 days of the state transition outlined by \citet{shidatsu2019} motivated by the clear drop in radio emission seen in \Cref{figure:radio_xray_correlation} before the hard to soft state transition. The points are encircled in \Cref{figure:radio_xray_correlation}.

\begin{figure}
\includegraphics[width=\columnwidth]{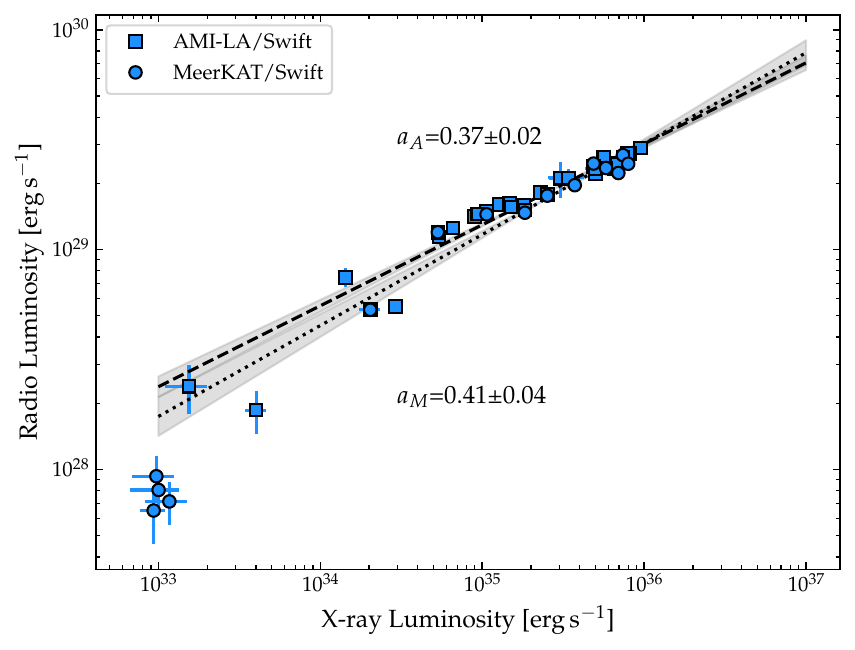}
\caption{As for \Cref{figure:radio_xray_correlation} but only including data post MJD 58500 which includes the hard-state-only re-brightenings only. Due to our lack of early MeerKAT coverage the fit for those data is similar to the one in \Cref{figure:radio_xray_correlation} at $a_{M}=0.41\pm0.04$, whereas the AMI-LA correlation is significantly shallower with $a_{A}=0.37\pm0.02$.}
\label{figure:radio_xray_correlation2}
\end{figure}

To check for different behaviour during the three re-brightening events we show the radio X-ray correlation for data post MJD 58500 in \Cref{figure:radio_xray_correlation2}. The slope during the re-brightenings is best measured by the AMI-LA (squares) and shows an apparent flattening from $L_{\rm{R}}\propto L_{\rm{X}}^{0.481\pm0.009}$ to $L_{\rm{R}}\propto L_{\rm{X}}^{0.37\pm0.02}$. It is interesting to note that this is the same X-ray luminosity range that showed a flatter correlation in \citealt{bright2020}, which just considered R0. The slope derived from the MeerKAT data is consistent with the full correlation with $L_{\rm{R}}\propto L_{\rm{X}}^{0.41\pm0.04}$.

\subsubsection{Optical -- X-ray Correlation}
\begin{figure}
\includegraphics[width=\columnwidth]{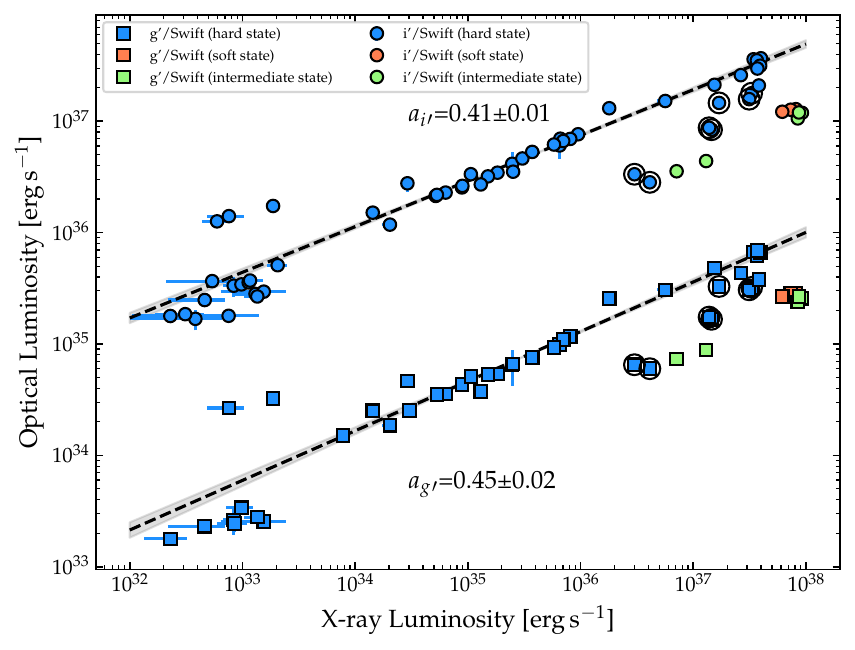}
\caption{The $i'$-band (top; multiplied by $10^{2}$ for clarity) and $g'$-band (bottom) optical -- X-ray correlation for J1820. Blue, green, and orange points are the hard, intermediate, and soft states, respectively. For the hard state we distinguish points before MJD 58233.5 by circling them and ignoring them when fitting the hard state correlation as they clearly do not follow the X-ray emission as seen for the rest of the hard state, falling significantly below the correlation at the end of the rising hard state. The black dashed lines shows a power-law fit to the blue data (minus those circled) with indices $a=0.45\pm0.02$ and $a=0.41\pm0.01$ for $g'$- and $i'$-band, respectively.\label{figure:optical_X-ray_correlation_plot}}
\end{figure}

The optical -- X-ray correlation is shown for $g'$- and $i'$-band data in \Cref{figure:optical_X-ray_correlation_plot}. During the majority of the hard state there is a clear correlation between optical and X-ray frequencies, however there are clear deviations from the general correlation at low and high luminosities. Indeed, it is clear from \Cref{fig:mw_light_curves} that during the initial hard state (also known as the rising hard state, before $\sim$ MJD 58300), after the first peak, the optical emission was dropping while the X-rays and radio stayed relatively constant (and correlated, see \Cref{figure:radio_xray_correlation}) before dipping slightly then rising as the X-ray spectrum softens and the hard to soft state transition occurred (the optical fade and dip, and the dependence of the optical--X-ray correlation with X-ray energy, are investigated in \citealt{yang2025}). We circle these points in the optical -- X-ray correlation in \Cref{figure:optical_X-ray_correlation_plot} and do not include them when fitting a power law to the relationship, which we find to be $L_{g\rm{'}}\propto L_{\rm{X}}^{0.45\pm0.02}$ and $L_{i\rm{'}}\propto L_{\rm{X}}^{0.41\pm0.01}$ in the hard state. These are similar to the optical--X-ray correlation slopes of J1820 measured in the soft state, $L_{O}\propto L_{X}^{0.51\pm0.03}$, by \cite{shidatsu2019}, but significantly shallower than the (population based) $L_{\rm{OIR}}\propto L_{\rm{X}}^{0.61\pm0.03}$ measured in \citep{russell2006} (although the correlation was constructed using optical/near-IR bands B to K). It is also clear that the correlation has more scatter at low luminosities, at around $10^{33}\,\rm{erg}\,\rm{s}^{-1}$ in X-ray luminosity, where the optical is over-predicted, but also with a few optical points brighter than the extrapolated correlation. These high luminosity optical points correspond to the soft to hard state transition and are discussed in detail later \Cref{sec:discussion_morph}.


\subsubsection{Radio -- Optical Correlation}
\begin{figure}
\includegraphics[width=\columnwidth]{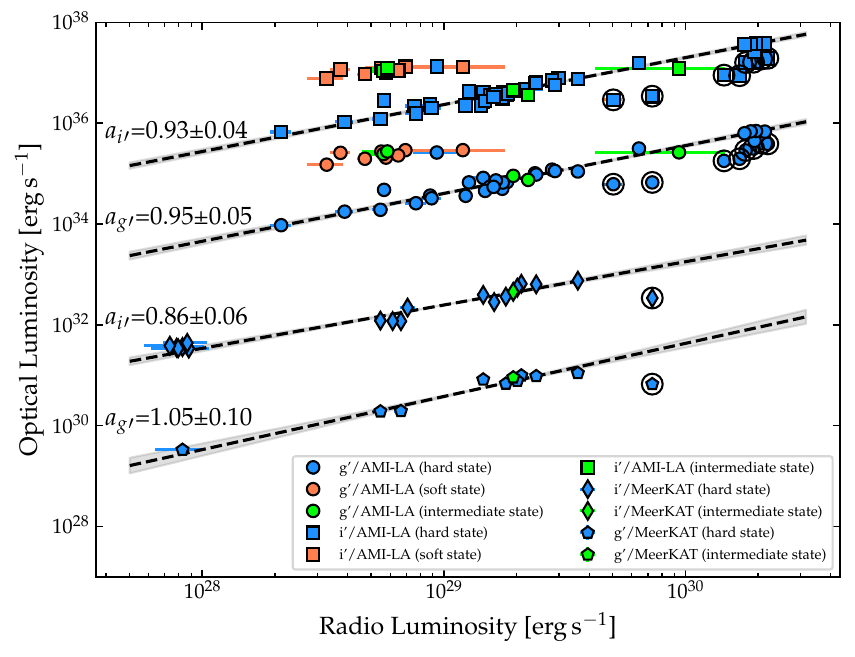}
\caption{The radio -- optical correlation for J1820 using data from MeerKAT, AMI-LA, and the $g$- and $i$- filters from LCO. The power law index of the fit for each pair is given next to the corresponding line. Blue, green, and orange points are the hard, intermediate, and soft states, respectively. From top to bottom the correlations shows optical $i'$ with AMI-LA (squares), optical $g'$ with AMI-LA (circles), optical $i'$ with MeerKAT (diamonds), and optical $g'$ with MeerKAT (pentagons). The optical luminosities have been offset to allow multiple correlations to be displayed on the same figure. For the hard state we distinguish points before MJD 58233.5 by circling them and ignoring them when fitting the hard state correlation. Data have been offset by $10^2$, $10^{0}$, $10^{-2}$, and $10^{-4}$ from top to bottom for clarity.}
\label{figure:radio_optical_correlation}
\end{figure}

Similar to the radio -- X-ray and optical -- X-ray correlations there is evidently strong coupling between radio and optical wavelengths. \Cref{figure:radio_optical_correlation} demonstrates these correlations using both of our optical bands and radio data at $1.28\,\rm{GHz}$ and $15.5\,\rm{GHz}$ from MeerKAT and the AMI-LA, respectively. As seen for the optical -- X-ray correlation a drop in the optical without response from the radio emission is evident during the rising hard state, followed by a rapid drop in the radio during the hard to soft state transition with a marginal re-brightening in the optical (most clear when correlating with the AMI-LA data due to the higher temporal density of the sampling). As before, we ignore data before MJD 58233.5 when fitting the correlations, which we find to be $L_{i\rm{'}}\propto L_{\rm{R}}^{0.93\pm0.04}$, $L_{g\rm{'}}\propto L_{\rm{R}}^{0.95\pm0.05}$, $L_{i\rm{'}}\propto L_{\rm{R}}^{0.86\pm0.06}$, and $L_{g\rm{'}}\propto L_{\rm{R}}^{1.1\pm0.1}$ from top to bottom in \Cref{figure:radio_optical_correlation}. The top two correlations are made with AMI-LA data and the bottom two with MeerKAT data. The measured correlation indices are broadly consistent, with the two MeerKAT -- optical correlations less well-constrained than those from the AMI-LA. During the intermediate state the radio quenches significantly with little response in the optical.\\

We show the fitting results of all two dimensional correlations in \Cref{tab:correlations}, including the correlation index, scaling factor, and the reduced $\chi$-square statistic for each fit. The reduced $\chi$-squared values for all of our fits are significantly larger than one, indicating residual scatter not being described by the model. This is likely caused by our data not being strictly simultaneous, some uncorrelated optical -- X-ray behaviour in the system, or short term optical variability \citep[e.g.][]{paice2019}.

\subsubsection{Radio -- X-ray -- Optical Correlation}

\begin{figure*}
\includegraphics[width=\textwidth]{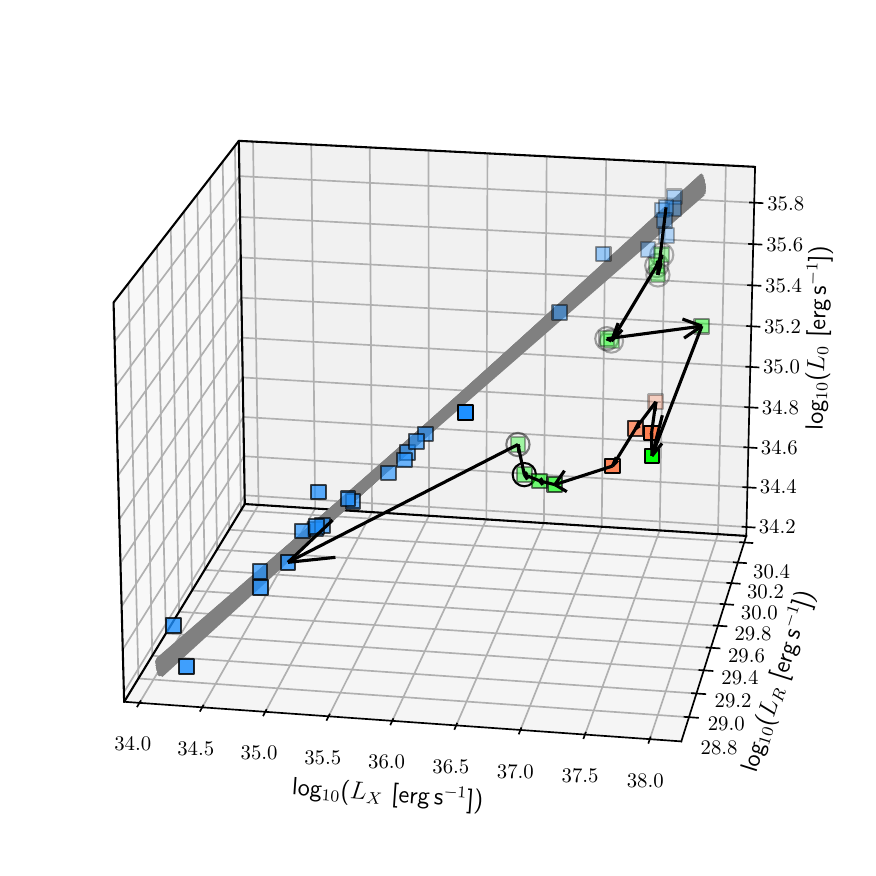}
\caption{The radio -- X-ray -- optical correlation plane for J1820. Blue, green, and orange points indicate that the source was in the hard, intermediate and soft states, respectively. The grey shaded region indicates the best fitting line to the hard state data from all three wavelengths with the thickness of the region showing the 1$\sigma$ error on the fit. Errors on data points are not shown on this plot. The black solid arrows connect points chronologically from the final point during the rising hard state, through the intermediate and soft states, and back to the first point of the fading hard state to guide the eye. The transparency of point indicates their depth in the three dimensional space, with fainter points at a greater depth. See \url{https://joesbright.github.io/MAXIJ180} for a fully interactive version of this figure.}
\label{figure:3d_correlation}
\end{figure*}

We show a projection of the three dimensional radio -- X-ray -- optical activity space of J1820 throughout its outburst in \Cref{figure:3d_correlation}. Unsurprisingly, during the hard state, the three wavebands are strongly correlated and lay along a line with centroid $[L_{\rm{X},0}, L_{\rm{R},0}, L_{\rm{O},0}] = [35.984\pm0.003, 29.510\pm0.006, 35.105\pm0.008]$ and unit direction vector $[a, b, c] = [0.846\pm0.002, 0.383\pm0.003, 0.371\pm0.005]$, with parameters defined in \Cref{sec:correlations}. The dropping optical emission in the rising hard state can be seen in the cluster of blue points at highest optical luminosity, and a hysteresis pattern through the hard -- soft -- hard transition series is evident. Due to the requirement of having three data points within a time bin we have fewer points in the three dimensional correlation when compared to the two dimensional ones.

\begin{table}
\centering
\caption{Power-law fitting parameters between our radio, X-ray, and optical data when J1820 was in the hard accretion state. Data are fit according to $L_{1}=AL_{2}^{a}$ (for $L_{1}$ -- $L_{2}$ in the table below), where $a$ is the power-law index and $A$ is the scaling factor. The reduced $\chi$-squared statistic ($\chi$-squared per degree of freedom) $\chi_{\nu}$ is given in the final column which describes the quality of the fit.}
\label{tab:correlations}
\begin{tabularx}{\columnwidth}{cccc}
\hline
Correlation & Power-law index ($a$) & Scaling factor ($A$) & $\chi_{\nu}$\\
\hline
$1.28\,\rm{GHz}$ -- X-ray & $0.41\pm0.03$& $15\pm1$ & 9.4\\
$15.5\,\rm{GHz}$ -- X-ray & $0.481\pm0.009$ & $12.2\pm0.3$ & 11.3\\
$g'$-band -- $1.28\,\rm{GHz}$ & $1.1\pm0.1$ & $4\pm3$ & 3.9\\
$i'$-band -- $1.28\,\rm{GHz}$ & $0.86\pm0.06$ & $9\pm2$ & 4.2\\ 
$g'$-band -- $15.5\,\rm{GHz}$ & $0.95\pm0.05$ & $7\pm1$ & 6.8\\
$i'$-band -- $15.5\,\rm{GHz}$ & $0.93\pm0.04$ & $7\pm1$ & 4.5\\
$g'$-band -- X-ray & $0.45\pm0.02$ & $19.1\pm0.7$ & 12.4\\
$i'$-band -- X-ray & $0.41\pm0.01$ & $20.1\pm0.5$ & 7.7\\
\hline
\end{tabularx}
\end{table}

\begin{figure*}
        \centering
        \begin{subfigure}[b]{0.49\textwidth}
            \centering
            \includegraphics[width=\textwidth]{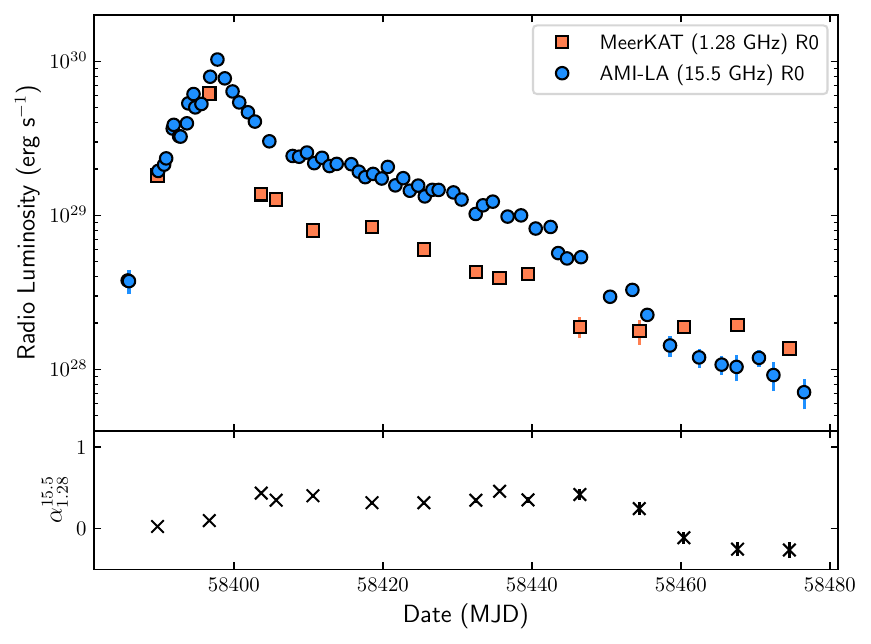}
            \caption[]%
            {{\small The hard state after J1820 exited the soft state, R0.}}    
        \end{subfigure}
        \hfill
        \begin{subfigure}[b]{0.49\textwidth}  
            \centering 
            \includegraphics[width=\textwidth]{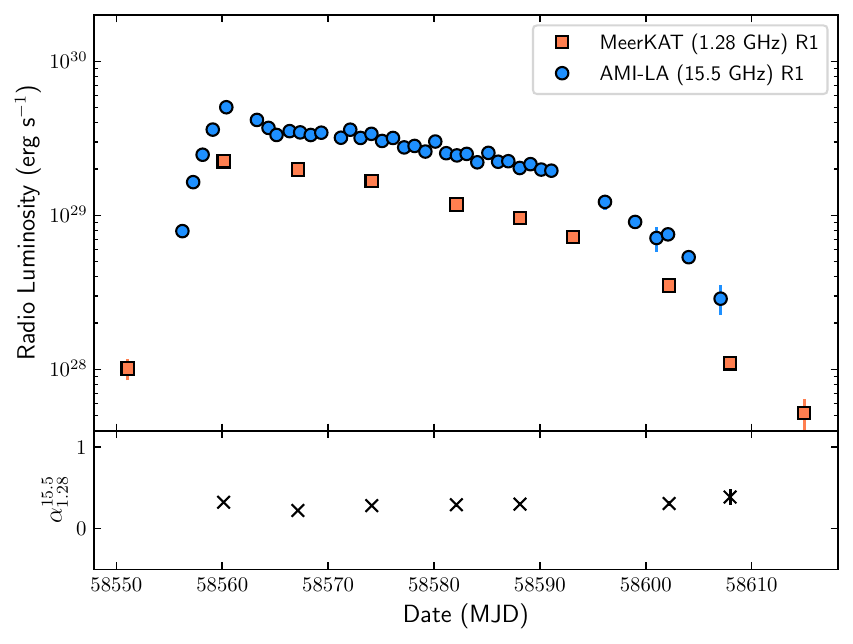}
            \caption[]%
            {{\small The hard-state only re-brightening event R1.}}    
        \end{subfigure}
        \vskip\baselineskip
        \begin{subfigure}[b]{0.49\textwidth}   
            \centering 
            \includegraphics[width=\textwidth]{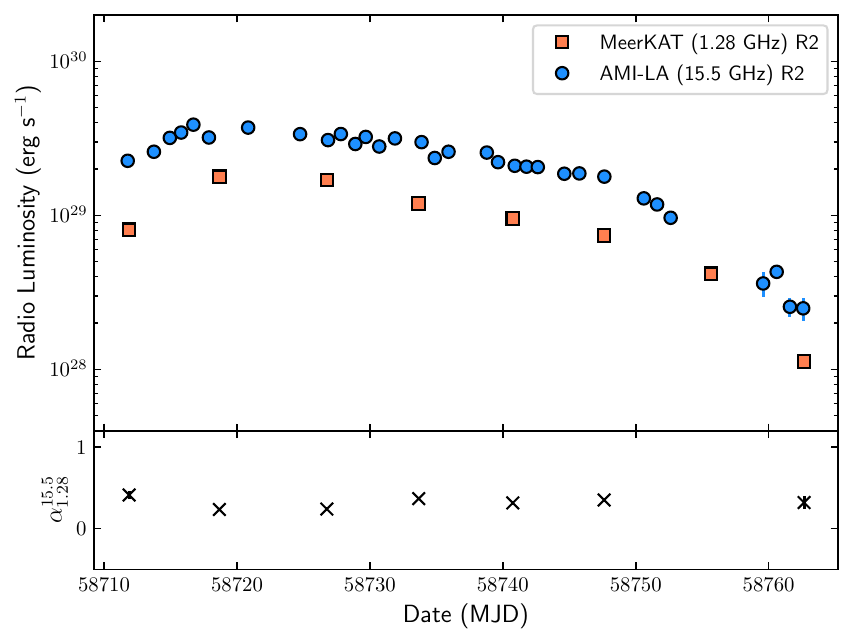}
            \caption[]%
            {{\small The hard-state only re-brightening event R2.}}    
        \end{subfigure}
        \hfill
        \begin{subfigure}[b]{0.49\textwidth}   
            \centering 
            \includegraphics[width=\textwidth]{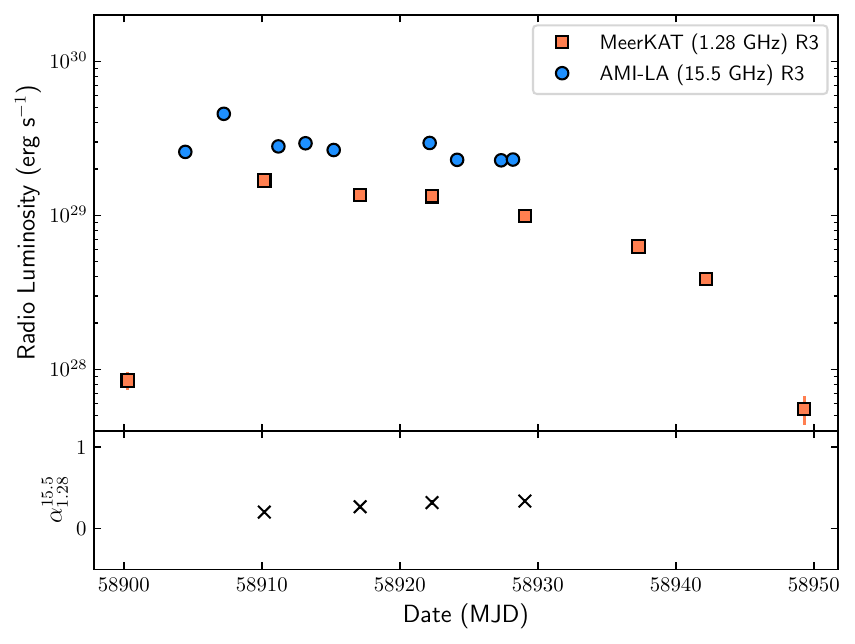}
            \caption[]%
            {{\small The hard-state only re-brightening event R3.}}    
        \end{subfigure}
        \caption[]
        {\small Light curves and the 1.28 to $15.5\,\rm{GHz}$ spectral index of the three re-brightening events R1, R2, and R3 (b, c, and d, respectively). Also shown is the hard state that J1820 entered immediately after its only excursion to the soft state (R0; panel a). Note that all plots share the same radio luminosity scale to allow for easier comparison between events.} 
        \label{fig:re-brightenings}
    \end{figure*}

\section{Discussion}\label{sec:discussion}
\subsection{Light curve morphologies}\label{sec:discussion_morph}
The radio, X-ray, and optical light curves of J1820 (\Cref{fig:mw_light_curves}) represent a complete look at the entirety of a canonical BHXRB outburst and multiple hard-state-only re-brightening events. There are a number of striking features that we highlight here (and which will be discussed in more detail in the following subsections). The most obvious is the high degree of correlation in the three observing bands. During both hard states in the canonical outburst (the rising hard state and R0) the three bands evolve in a similar way, with the exception of the significant jet quenching in the soft state indicated by the sharp drop in radio flux density  \citep[with the low level emission caused by transient ejecta, see][]{bright2020,espinasse2020,wood2021} and the early fading optical in the rising hard state. The similarity of the multi-wavelength data during the three re-brightening events is particularly notable (see \Cref{fig:re-brightenings}), with the same sharp rise to maximum, and changing decay rate during the decay, visible in all bands. The peak of the three re-brightenings is also relatively stable, as is their spacing. There is also an obvious similarity with the fading hard state (R0; which becomes optically thin around MJD 58460, likely due to the ejecta becoming dominant) and the re-brightening events (R1, R2, R3) not seen in the rising hard state. Optical observations during the periods between re-brightenings show a persistent level of emission also seen at X-ray and radio wavelengths, indicating that the source did not reach quiescence between re-brightenings \citep{baglio2023atel}. The optical, X-ray, and radio observations made after MJD $\sim59000$ indicate that J1820 has entered a relatively stable phase with a slowly rising optical (and possibly X-ray) luminosity. While our radio and X-ray coverage was less complete in this phase it is unlikely that we missed any similar re-brightening events, given that they would have been detectable by our comparatively more regular optical monitoring since R3. Note that optical observations taken in June 2023 (MJD 60106; later than the observations presented in \Cref{fig:mw_light_curves}) suggest that the optical emission from J1820 finally reached its pre-outburst level after 5 years of activity \citep{baglio2023atel}.

We note the sharp rise in optical luminosity after the return to the hard state at MJD $\sim58400$ while the radio emission rises rapidly and the X-ray emission is declining (see \Cref{fig:compact_formation}). If the optical emission we see is entirely associated with reprocessed X-ray emission from the inner accretion disk then we would expect the optical emission to follow it tightly. This opposite behaviour is suggestive of a potential jet contribution to the optical, with the radio and optical emission now coupled instead. This jet contribution to the optical emission has been confirmed by broadband spectral modelling \citep{ozbey2022,trujillo2024} and optical polarisation evolution during state transitions \citep{veledina2019}.

\begin{figure}
\includegraphics[width=\columnwidth]{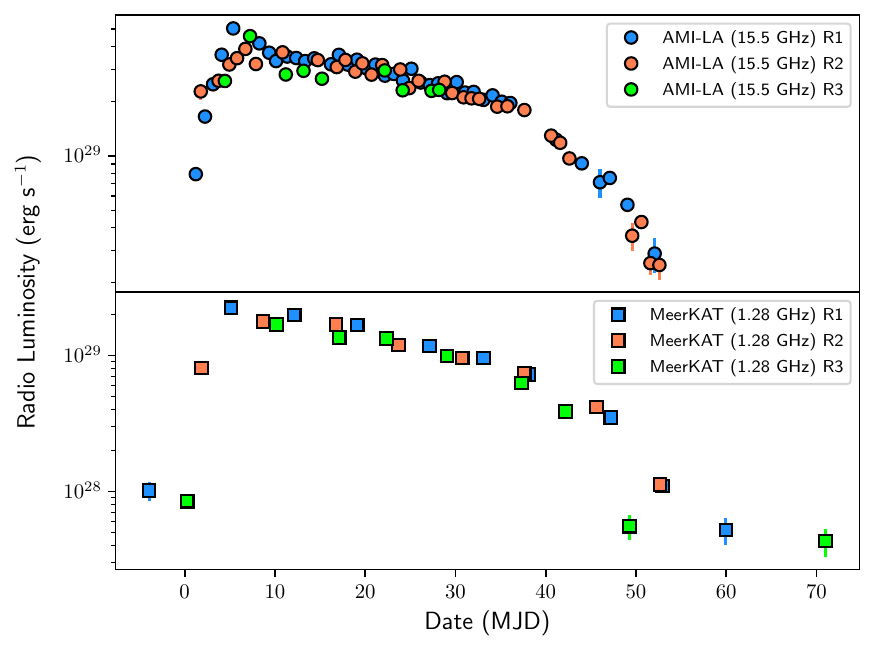}
\caption{The three hard-state-only re-brightenings shown by J1820 (blue, red, and green, chronologically), as seen at $1.28\,\rm{GHz}$ (squares) at $15.5\,\rm{GHz}$ (circles). The re-brightenings have been shifted in time to overlap the first re-brightening in order to demonstrate their regularity. The second and third re-brightenings have been shifted by $155\,\rm{days}$ and $345\,\rm{days}$, respectively. Denoting the start time of rebrightening episode RX as $t_{\rm{RX}}$ these offsets imply $t_{\rm{R2}}-t_{\rm{R1}}=155\,\rm{days}$ and $t_{\rm{R3}}-t_{\rm{R2}}=190\,\rm{days}$.}
\label{fig:stacked_re-brightenings}
\end{figure}

The re-brightening events R1, R2, and R3 show remarkable similarities in their light curve morphologies as seen from both MeerKAT and the AMI-LA (\Cref{fig:re-brightenings,fig:stacked_re-brightenings} show the radio re-brightenings). It is also notable that the re-brightening events are relatively regularly spaced and reach a similar peak flux density for each flare. The hard state at the end of the canonical outburst (R0) also shows similarity to the re-brightening events with the exception of a higher peak flux density. The hard state immediately preceding the soft state shows no such similarity, demonstrating a relatively constant flux for around 100 days after the start of the outburst. The regularity of the re-brightenings (both temporally and in morphology) suggests that their evolution is being regulated by a common process. The peak X-ray luminosity of the re-brightenings, $\sim10^{36}\,\rm{erg}\,\rm{s}^{-1}$, corresponds to $\sim0.2$ per cent of the Eddington limit for a black hole mass of $8.5\,M_{\odot}$ \citep{munoz-darias2019} and the total flare timescale is around 50 to 60 days. The spectral index evolution throughout each re-brightening is also similar, showing a flat or slightly inverted radio spectrum for the duration. There are also two clear decay rates during the fading, with a shallow decay for around 30 days followed by a steeper decay for around 20 days until radio emission is no longer detected. This is most clearly seen in the AMI-LA radio and optical light curves due to their higher time density sampling.

\cite{saikia2023a} compiled a list of LMXBs with re-brightening events within one year of the end of the initial outburst. Similarities are clear between J1820 and the optical observations of Swift J1910.2$-$0546 (J1910.2) which showed optical re-flaring on a $\sim45$ day timescale, although with less well-sampled light curves than for J1820. The flaring in J1910.2 also peaked at an approximately consistent luminosity, similar to J1820, although the spacing between flares was significantly shorter, and the flare profile different. The lack of comprehensive X-ray and radio observations prevents comparison at these wavelengths. Re-flaring events have also been observed in MAXI J1535$-$571 \citep[J1535][]{parikh2019,cuneo2020}, however in this case the flare amplitude decreased with each subsequent flare and the profile was markedly smoother and more symmetric than for J1820. Additionally J1535 was undergoing state transitions (seen in the hardness intensity diagram and through jet quenching) which J1820 was certainly not \citep{parikh2019}. The similarity between the flaring in J1820 and J1910.2 suggests that J1910.2 did not undergo state transitions during its re-flaring, a conclusion also reached by \cite{saikia2023b}. The late time radio increases from MAXI J1348$-$630 appear to have roughly the same spacing as for J1820, and while they are less well-sampled their shapes are consistent \citep{carotenuto2021a}. Finally, \cite{chen1997} present a large sample of X-ray light curves from X-ray novae, with the source XN 0422$+$32 showing similar flaring to J1820.

In the case of J1820 the similarity of R1/R2/R3 to R0 indicates that the morphology of the re-brightening events can occur during the canonical outburst. The re-brightening events themselves show some similarity to those predicted from disk instability models where irradiation heating from the inner accretion disk (or corona) is included \citep{dubus2001}. In this model the length of the flare is set by the strength of the irradiation and the disk mass transfer rate, whereas the profile is set by an initial exponential decay where the hot disk is kept fully ionised and no cooling front can form, followed by a cooling front that moves inwards due to weakening irradiation. Finally, once the irradiation level drops significantly enough, the evolution continues on the thermal timescale \citep{dubus2001, btetarenko2018b}. From an observational perspective this should result in a disk light curve profile described by a sharp rise, followed by a decaying phase which switches to a steeper day during the decline. This is morphologically consistent with the light curves shown in \Cref{fig:re-brightenings}, with a slow linear (in the log -- linear plot) decay followed by a more rapid decline. Even though these radio light curves are probing the core jet, a similar profile is seen in the radio and optical.

Finally, we check for evidence of changes in the X-ray -- radio correlation during the evolution of the re-brightening events. Despite the clear change in decay rate seen at $\sim40$ days in \Cref{fig:stacked_re-brightenings}, we see no indication that the correlation index evolves throughout the re-brightenings, which could have indicated a change in the accretion efficiency (e.g. to an advection dominated accretion flow) during the evolution.

\subsection{Formation of the compact jet}
\begin{figure}
\includegraphics[width=\columnwidth]{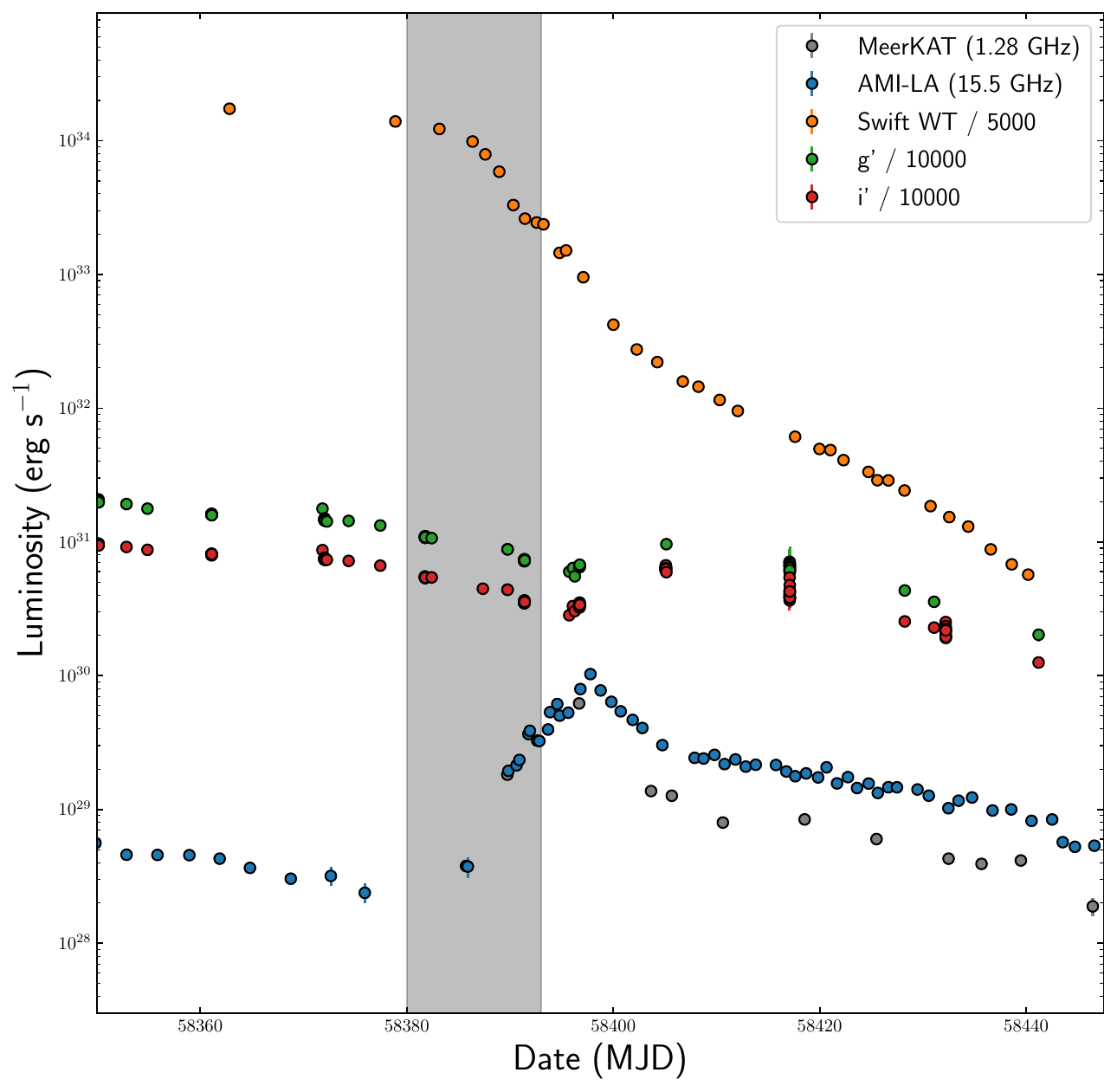}
\caption{The soft to hard state transition showing the re-ignition of the core jet, particularly obvious at radio frequencies. The X-ray and optical luminosities have been scaled down by a factor of 5000 and 10000, respectively, to ease comparison.}
\label{fig:compact_formation}
\end{figure}

In \Cref{fig:compact_formation} we show the transition from the soft to hard state for J1820 with data at radio, optical, and X-ray observing bands. We see an initial hardening of the radio spectrum as the compact core jet re-forms during the intermediate state, while the soft X-rays fade and the optical rises at the end of the intermediate state. The formation of the compact jet is seen in similar detail in \citet{corbel2013} for GX339$-$4 (see also \citealt{coriat2009}), where the evolution qualitatively matches J1820. It has been suggested that the ordering of the radio/optical rising (which is seen in reverse in the hard to soft state transition) is due to an evolution of the compact/core jet structure where different regions become optically thick on different timescales \citep[e.g.][]{millerjones2012,russell2013,corbel2013,kalemci2013,russell2014}. This could also be responsible for the drop in optical luminosity that we see during the initial hard state which is evident in \Cref{fig:mw_light_curves,figure:radio_optical_correlation}, with the ordering reversed (the region of the jet closest to the black hole quenching before those at larger radii \citealt{coriat2009,yan2012,russell2020}).

\subsection{Jet contribution to optical emission}
It is well known that during the hard to soft state transition in BHXRBs the core jet is significantly quenched \citep{fender1999_quench,russell2019,bright2020} or switches off entirely. This is most readily seen in high angular resolution and sensitivity radio observations, which can distinguish any contamination from the launch of transient ejecta \citep{bright2020,wood2021}. The state transition is also associated with a change in the configuration of the inner accretion disk, responsible for the X-ray emission, with the peak of the X-ray shifting to $\sim1\,\rm{keV}$. This can be seen in \Cref{fig:mw_light_curves} where the soft X-ray luminosity as measured by Swift rises as J1820 enters the soft state. From \Cref{figure:radio_optical_correlation} we see a clear signature of optical quenching before the source entered the hard to soft intermediate state. The quenching is significant, about 1 dex, and precedes a large drop in the radio flux density. The drop from MJD $\sim58200$ to MJD $\sim58230$ is likely to be at least partly caused by the optical emission from the jet quenching before the radio as the spectrum `collapses' from higher to lower frequencies. In this scenario the remaining optical emission is therefore produced by remaining sources of optical emission, most likely dominated by the accretion disk (both directly, and through reprocessing). This is consistent with the broadband SED modelling performed in \citealt{trujillo2024} (The optical fading is also discussed in the context of the optical/X-ray correlation in \citealt{yang2025}).\\

The expected correlation between optical and radio emission depends on the emission mechanism primarily responsible for the production of optical photons. While radio emission from hard state BHXRBs is well established to be produced in a compact jet, the optical emission is likely some combination of outer disk photons, reprocessed inner disk photons, and photons from a region of the jet closer to the black hole than the ones responsible for the radio emission \citep[the companion star is sub-dominant, see e.g.][]{trujillo2024}. \cite{russell2006} showed that the optical -- X-ray correlation in the soft state was mostly unchanged when compared to the hard state, whereas the infrared was significantly suppressed compared to the X-rays, indicating that the jet is contributing near infrared (NIR) photons which are quenched along with the radio in the soft state. It is likely that any optical jet emission is above the jet break frequency where the jet synchrotron emission becomes optically thin, as shown by \cite{trujillo2024} for J1820, and so the relative contribution will be suppressed when compared to NIR and therefore the impact on the light curves more subtle. 

In the case that the optical emission is primarily caused by the reprocessing of inner disk X-ray photons, a theoretical correlation between the two components of $L_{\rm{O}}\propto L_{\rm{X}}^{0.5}a$ is expected \citep{vanparadijs1994}, where $a$ is the orbital separation of the system (which remains constant for well circularised binaries). Coupling this with the radio X-ray correlation of $L_{\rm{R}}\propto L_{\rm{X}}^{\sim0.6}$, well established for radio-loud black hole X-ray binaries in the hard state (and consistent with originating from a radiatively inefficient accretion flow, e.g. \citealt{narayan1995,kording2006}), gives the prediction $L_{\rm{O}}\propto L_{\rm{R}}^{\sim0.8}$, entirely consistent with the correlations shown in \Cref{figure:radio_optical_correlation} for J1820 when adjusting for the shallower radio -- X-ray correlation found for this source. Note that this correlation holds well using radio measurements an order of magnitude different in frequency, and for multiple optical bands (although the errors on the MeerKAT -- optical observation are significantly larger than for the AMI-LA -- optical correlation). 

If instead optical emission is produced primarily in the jet, and the optical and radio emission are both produced as part of the same power-law components on the spectrum (e.g. the flat self-absorbed core spectrum\footnote{Or even if the optical is in the optically thin synchrotron regime, since the jet spectral break does not shift dramatically in frequency at different luminosities in the hard state \citep{Russell2013jetbreaks}.}) then, of course, $L_{\rm{O}}\propto L_{\rm{R}}$ is expected, with the same correlation between X-ray and optical emission as between X-ray and radio emission. This scenario is indistinguishable from the reprocessing model due to the flatter radio -- X-ray correlation for J1820 and so based on correlation arguments no conclusions can be drawn on the presence of optical photons. We note that the optical -- X-ray correlations shown in \Cref{figure:optical_X-ray_correlation_plot} have a marginally shallower index than expected for the model of \cite{vanparadijs1994}.

\subsection{The Radio -- X-ray Correlation}
The radio -- X-ray correlation is a widely used diagnostic for accreting systems, probing the connection between accretion and the process of jet production and propagation. Active galactic nuclei (AGN) and a subset of the X-ray binaries lie on the fundamental plane of black hole activity, showing $L_{\rm{R}}\propto L_{\rm{X}}^{0.6}$ when marginalising over the black hole mass \citep{merloni2003,falcke2004}. The observations presented in this work, and in \cite{bright2020}, mark J1820 as one of the best sampled sources in the radio -- X-ray plane, second only to GX 339$-$4 \citep[e.g.][]{corbel2003,tremou2020}. The best correlation we find for J1820 of $a=0.482\pm0.007$ is shallower than for the fundamental plane of black hole activity, however J1820 is clearly similar to the radio loud sources for the entirety of its outburst in terms of radio luminosity. Sharp deviations from the radio -- X-ray correlation can be seen when the source transitions between the hard and soft accretion states, with a return to the correlation from the reverse transition. This behaviour tracks the quenching and re-ignition of the core jet. Radio detections in the soft state have been unambiguously associated with the presence of transient ejecta no longer connected to the accretion processes close to the black hole \citep{bright2020,espinasse2020,wood2021}.

The high density of our sampling at X-ray and radio wavelengths allows us to test for possible changes in the correlation over the outburst. It has been suggested \citep{coriat2011} that the correlation index can be directly associated with the efficiency of accretion onto the compact object, with the radio loud and radio quiet corresponding to inefficient and efficient accretion states, respectively. Radio quiet sources rejoin the loud track at low luminosities, possibly indicating a switch in accretion efficiency. Radio loud sources seem to never deviate from the track, although the correlation for J1820 is shallower than the canonical slope. For J1820 there appears to be a flattening in the radio X-ray correlation to $a=0.35\pm0.02$ when the X-ray luminosity is between $\sim10^{35}$ and $\sim10^{36}\,\rm{erg}\,\rm{s}^{-1}$ for both the fading hard state and the re-brightenings, occurring in a similar luminosity range to that seen for radio quiet sources as they rejoin the radio loud track \citep[e.g.][]{carotenuto2021b}. It has also been suggested that a shallower radio -- X-ray correlation is a feature of hard-state only outbursts from GX 339$-$4, which \citet{dehaas2021} suggest could be driven by a different coupling between the jet and the accretion flow in these outbursts. It should be noted that while individual sources clearly traverse different paths in the radio -- X-ray plane, the existence of two distinct tracks is disputed, as is any difference in the correlation index for BHs and neutron stars \citep{gallo2018}. Any such difference would be an important consideration for models attempting to explain the radio -- X-ray correlation using accretion efficiency \citep[see e.g.][]{gusinskaia2020,gasealahwe2023}.

\subsection{A three dimensional activity plane}

The creation of a three dimensional radio -- optical -- X-ray activity plane for J1820 provides a full picture of its evolution. It is clear that during the hard state the three wavelengths are tightly coupled (unsurprisingly, given \Cref{fig:mw_light_curves}), apart from during the rising hard state where the optical faded independently. The hysteresis pattern regularly seen in the radio -- X-ray correlation plane is present also in three dimensions, with the optical rejoining the radio -- optical --X-ray correlation at a lower luminosity. The fading hard state and three re-brightening events are indistinguishable in this space. A fully interactive version of this figure is available at \url{https://joesbright.github.io/MAXIJ180}.

\section{Conclusions}\label{section:conclusions}
We present high temporal density radio, X-ray and optical monitoring observations of the black hole X-ray binary MAXI J1820 over more than 4 years (from 12 March 2018 to 1 August 2022), beginning with its first recorded outburst and including multiple re-brightening events. These data are among the most comprehensive that exist for a BHXRB at these wavelengths. We have constructed all possible correlations between these three observing bands, including the three dimensional activity plane, providing new insights into the evolution of XRB outbursts. Particularly, we see that during the rising hard state the radio and X-ray emission are strongly coupled and relatively stable, while the optical emission decays significantly. We speculate that this could be the result of the compact jet quenching from its base and then gradually to larger size scales down the flow. During the soft to hard state transition we see peaks in the radio and optical emission ordered in reverse when compared to the hard to soft transition. In the fading hard state, and subsequent re-brightenings, the optical emission is well coupled to the X-ray and radio emission at all times. Based on our correlations we are not able to determine the source of the optical jet photons due to a similarity in the predictions for the relationships between optical emission from an irradiated accretion disk and from a flat spectrum jet.

The regularity of the re-brightening events (and to a lesser extent the fading hard state) and their structured morphology bears similarities to those predicted by extensions to the disk instability model that include irradiation heating from the central accretion disk. Particularly, we see an exponential decay followed by a sharper decline, which are controlled by the viscous timescale and cooling timescales, respectively \citep{dubus2001,btetarenko2018a,btetarenko2018b}. 

For sources with appropriately sampled data the formation of a radio -- optical -- X-ray plane allows for a better understanding of the complex interplay between accretion and jet production. This is particularly important for the optical emission, which has contributions from both the accretion disk and the jet. In the near future the formation of the radio -- optical -- X-ray plane should be possible for the bright BHXRB Swift J1727.8$-$1613, which has exceptional radio coverage \citep{hughes2025}.

\section*{Acknowledgements}
We thank Dan Bramich for the development of the XB-NEWS pipeline used for extracting the optical LCO magnitudes.
RF acknowledges support from UKRI, The European Research Council, and the Hintze Charitable Foundation.
JvdE is supported by funding from the European Union's Horizon Europe research and innovation programme under the Marie Skłodowska-Curie grant agreement No 101148693 (MeerSHOCKS), and acknowledges a Warwick Astrophysics prize post-doctoral fellowship made possible thanks to a generous philanthropic donation and a Lee Hysan Junior Research Fellowship awarded by St. Hilda’s College, Oxford.
We thank the staff at the Mullard Radio Astronomy Observatory for carrying out observations with the AMI-LA.
The MeerKAT telescope is operated by the South African Radio Astronomy Observatory (SARAO), which is a facility of the National Research Foundation, an agency of the Department of Science and Innovation. We thank the SARAO staff for conducting these observations.
We acknowledge the use of the ilifu cloud computing facility -- \url{www.ilifu.ac.za}, a partnership between the University of Cape Town, the University of the Western Cape, the University of Stellenbosch, Sol Plaatje University, the Cape Peninsula University of Technology and the South African Radio Astronomy Observatory. The ilifu facility is supported by contributions from the Inter-University Institute for Data Intensive Astronomy (IDIA -- a partnership between the University of Cape Town, the University of Pretoria and the University of the Western Cape), the Computational Biology division at UCT and the Data Intensive Research Initiative of South Africa (DIRISA).
This work makes use of observations from the Las Cumbres Observatory global telescope network.
We acknowledge the use of public data from the Swift data archive. This work made use of data supplied by the UK Swift Science Data Centre at the University of Leicester.
This material is based upon work supported by Tamkeen under the NYU Abu Dhabi Research Institute grant CASS.
This work benefited from discussions during Team Meetings of the International Space Science Institute (Bern), whose support we acknowledge.
We thank the anonymous referee for their comments, which helped improve this manuscript.

\section*{Data Availability}
All data presented in this work are provided in machine readable format as part of the online version of the paper. Radio and optical images from which flux densities were derived are available upon reasonable request to the corresponding author. \textit{Swift} data products can be recreated from the public data archive and using Xspec in HEASoft.



\bibliographystyle{mnras}
\bibliography{1820} 




\bsp	
\label{lastpage}
\end{document}